\documentclass[prb,amssymb,amsmath,reprint]{revtex4-1}
\usepackage{bm}
\usepackage{graphicx}
\usepackage{color}
\usepackage{amsmath}
\usepackage{amsfonts}

\newcommand{\g}{\mathcal{G}}
\newcommand{\T}{\mathcal{T }}

\newcommand{\vpht}{\vphantom{\dagger}}
\newcommand*{\cd}[2]{c^{\dagger}_{#1 {\vec k}_{#2}\sigma_{#2}}}

\newcommand*{\cv}[2]{c^{\vpht}_{#1 {\vec k}_{#2}\sigma_{#2}}}

\newcommand*{\ek}[1]{\epsilon^{#1}_{\vec k}}

\newcommand*{\sumk}[1]{\sum_{{\vec k}#1}}
\renewcommand*{\vec}[1]{\boldsymbol{#1}}

\begin{document}

\title{Magnetotransport in the Kondo model with ferromagnetic exchange
interaction}

\author{Mathias Cabrera Cano}

\affiliation{Institut N\'eel, CNRS and Universit\'e Joseph Fourier,
25 avenue des Martyrs, BP 166, 38042 Grenoble, France
}

\author{Serge Florens}%
\affiliation{Institut N\'eel, CNRS and Universit\'e Joseph Fourier,
25 avenue des Martyrs, BP 166, 38042 Grenoble, France
}

\date{\today}

\begin{abstract}
We consider the transport properties in an applied magnetic field of the spin
$S=1/2$ Kondo model with {\it ferromagnetic} exchange coupling to electronic
reservoirs, a description relevant for the strong coupling limit of
underscreened spin $S=1$ Kondo impurities. Because the ferromagnetic Kondo
interaction is marginally irrelevant, perturbative methods should prove accurate
down to low energies. For the purpose of this study, we use a combination 
of Majorana diagrammatic theory with Density Matrix Numerical Renormalization Group 
simulations. 
In the standard case of antiferromagnetic Kondo exchange, we first
show that our simple analytical technique recovers results for the $\T$-matrix
and spin relaxation at weak coupling (above the Kondo temperature), that
were previously obtained using functional renormalization tools. Considering
then the ferromagnetic case, we demonstrate how the low-energy Kondo anomaly
splits for arbitrary small values of the Zeeman energy, in contrast to fully
screened Kondo impurities near the strong coupling Fermi liquid fixed point, and
in agreement with recent experimental findings for spin $S=1$ molecular quantum
dots.

\end{abstract}

\maketitle

\section{Introduction}

Magnetotransport measurements in quantum dots allow to probe and tune
the magnetic properties of nanostructures in a situation where the
conductance is the only experimentally accessible
quantity~\cite{kastner_single-electron_1992,kastner_single_2000}. Cotunneling Zeeman 
spectroscopy under an applied voltage bias ({\it i.e.} in a process that 
does not electrically charge the quantum dot) can be used for instance to resolve 
the magnetic degeneracies of quantized energy levels, in similar spirit to 
usual optical analysis of atomic and molecular spectra [see
Refs.~\onlinecite{roch_quantum_2008},\onlinecite{Florens_Review} for an illustration in a two-electron
quantum dot showing nearly degenerate singlet and triplet states].
A striking difference between quantum electronic transport and quantum
optics is the easiness to realize the strong coupling of a discrete quantum
system to a macroscopic number of external degrees of freedom, simply by
contacting a single atom or molecule to metallic electrodes. 
While standard physical effects such as screening of the molecular charging 
energy~\cite{Flensberg_screening} and renormalizations of the atomic energy 
levels from quantum tunneling~\cite{Holm,Hauptmann,Florens_Review} can occur, these electronic 
setups opens a window to realize even more subtle many-body physics. One well-studied 
example is the Kondo effect in strongly tunnel-coupled quantum dots, where the
magnetic screening of a spin active discrete level is performed by the 
bath of surrounding conduction
electrons~\cite{hewson_kondo_1997,goldhaber-gordon_kondo_1998,Cronenwett,pustilnik_kondo_2004,GrobisReview}.
Interestingly, a variety of different and possibly exotic Kondo effects can occur 
in simple extensions of the Kondo model, several of which have been already realized 
experimentally [see Ref.~\onlinecite{Florens_Review} for a review]. 
While the standard Kondo effect involves a single spin $S=1/2$ coupled to a single 
electronic bath and leads to Fermi-liquid properties at low temperatures, a 
variant as simple as the spin $S=1$ case already presents anomalous transport 
properties~\cite{roch_observation_2009,ParksUS}. In the latter situation, conduction electrons cannot 
manage a full compensation of the magnetic impurity, resulting in a partially 
screened ground state with remanent $\log(2)$
entropy~\cite{nozieres_kondo_1980,Mehta,Koller,Logan}, 
yet unitary conductance at zero temperature $G=2e^2/h$ ($e$ is the electron 
charge and $h$ Planck's constant). At low energy, 
the remanent magnetic degrees of freedom decouple from the electronic reservoirs
(in a logarithmically slow fashion) and can be recovered by the application of
an arbitrarily small Zeeman energy~\cite{roch_observation_2009}. In contrast, the fully 
screened case of a spin $S=1/2$ leads to a singlet ground state with small but non-zero 
binding energy (the so-called Kondo temperature), which remains stable to the
application of a magnetic fields that do not exceed the Kondo energy. These simple 
examples illustrate nicely how Zeeman spectroscopy can be used to distinguish between 
different types of many-body states~\cite{roch_observation_2009,ParksUS}.

The purpose of this paper is to investigate in detail the magnetotransport properties
of such highly spin-polarizable underscreened $S=1$ Kondo impurities. We will make 
use here of the Nozi\`eres-Blandin description of the strong coupling fixed point
in terms of the spin $S=1/2$ Kondo model with {\it ferromagnetic} Kondo
exchange~\cite{nozieres_kondo_1980,Mehta,Koller}. In that case, the magnetic interaction between 
the quantum impurity and the reservoirs can be treated by perturbative methods, 
allowing a controlled description of the transport properties. While one could use 
here the functional renormalization group methodologies previously developed to deal 
with the weak coupling regime of the antiferromagnetic
case~\cite{rosch_spectral_2003,rosch_kondo_2005,Schoeller,Schuricht,SchurichtAndergassen,Kehrein}, we will show that 
a comparable description can be achieved by more standard diagrammatic methods, thanks 
to an useful mapping of a single spin $S=1/2$ in terms of Majorana
fermions~\cite{parcollet_perturbative_2002,shnirman_spin-spin_2003,mao_spin_2003,
bulaevskii_tunneling_2003,Florens_spinboson,Mathias_Kondo_Thesis}. The advantage of the Majorana technique 
with respect to more usual representations of the spin algebra lies in
the recent understanding that spin dynamics can be described by one-particle
Majorana Green's functions, allowing the use of Dyson's equation within the 
standard Feynman diagram expansion~\cite{shnirman_spin-spin_2003,mao_spin_2003,Florens_spinboson}. 
We will demonstrate that transport properties, as 
described by the $\T$-matrix of the Kondo problem, become similarly much simpler in the Majorana
language. Indeed, while standard spin representations would require the computation of
six-point correlation functions, the Majorana description expresses the $\T$-matrix
as a four-point (two-body) fermionic correlator, where the RPA expansion for 
response functions can be used at leading logarithmic order. 
In order to test the method, we will first discuss how the usual weak-coupling
description of the antiferromagnetic Kondo problem can be done within the
Majorana diagrammatics, both at zero and finite magnetic field. A comparison
to calculations based on the full density matrix extension~\cite{hofstetter_generalized_2000,toth_density_2008} 
of the Numerical Renormalization Group 
(NRG)~\cite{wilson_renormalization_1975,krishna-murthy_renormalization-group_1980,bulla_numerical_2008}, 
and previous analytical results~\cite{rosch_spectral_2003,rosch_kondo_2005} will be made. 
We will then compute in detail the spin and transport dynamics 
of a {\it ferromagnetically} coupled quantum dot, both within the Majorana description 
and the NRG simulations, which are shown to match quantitatively, apart from
a narrow domain near the Zeeman energy, where the NRG data lack sufficient
energy resolution. Our main
result concerns the Zeeman splitting of the $\T$-matrix in the ferromagnetic case,
which we show to occur for arbitrarily small values of the applied magnetic field, 
in agreement with recent experimental findings for underscreened $S=1$ Kondo impurities
in molecular quantum dots~\cite{roch_observation_2009}.
We also note that the ferromagnetic Kondo model considered here has also been
recently proposed for some types of atomic size metallic constrictions on the basis 
of ab-initio calculations~\cite{lucignano_kondo_2009}, but has received so far
little attention~\cite{Koller}.

The paper is organized as follows. In Sec.~\ref{sec:Kondo} we introduce
the Kondo model relevant for transport in quantum dots, discussing transport
and magnetic properties. The differences between  fully and partially screened moments,
with the connection to the ferromagnetic Kondo problem, will be presented.
In Sec.~\ref{sec:majo}, we develop the general Majorana diagrammatic method,
both at zero and finite magnetic field. Results are finally given in
Sec.~\ref{sec:results}, where both antiferromagnetically and ferromagnetically
coupled spin $S=1/2$ impurities are considered, using the Majorana technique
and NRG calculations. A critical discussion of numerical and analytical scheme
is given in the conclusion.

\section{Kondo models for transport in quantum dots}
\label{sec:Kondo}

\subsection{Spin ${\bm S=1/2}$ case}

The conductance $\g$ of a quantum dot is the main physical property that can be
directly measured in quantum dot experiments. In semiconducting dots, electrons
are confined to a small region with the help of top-gates, leading to a
set of discrete energy levels subject to electron-electron interaction
effects, and tunnel coupled to macroscopic reservoirs. Under
a voltage bias, a finite current flows through the dot,
and the conductance is determined by the transition rate of tunneling events from 
source to drain.
The concept of single-electron transistor comes from the capacitive
coupling in small confined dots, where the discrete number of 
electrons on the dot can be tuned using a 
backgate-voltage~\cite{kastner_single-electron_1992,kastner_single_2000}.
If the number of trapped electrons on the dot is odd, an electronic
configuration with an unpaired spin obtains, leading usually to a $S=1/2$ state.
The tunneling into the reservoirs then results in an effective {\it
antiferromagnetic} exchange interaction $J>0$, which tends to suppress 
the magnetic state of the dot. The physical origin of the antiferromagnetic exchange
is Pauli's principle, which favors second order tunnel processes for
antiparallel spin orientation of the electron in the dot and the ones in the
reservoirs. This interaction can be described by the Kondo Hamiltonian:
\begin{equation}
\label{eq:kondo_model}
H_{K}= J{\bf S}\;\cdot\!\!\sum_{\alpha_1 \alpha_2}
\sum_{ {\bf k}_1{\bf k}_2}\sum_{\sigma_1\sigma_2}
\cd{\alpha}{1}\frac{\vec{\tau}^{\vpht}_{\sigma_1\sigma_2}}{2}
\cv{\alpha}{2}
\end{equation}
where $\vec{\tau}$ denotes the vector $\vec{\tau}=\begin{pmatrix}\tau^x, &
\tau^y, & \tau^z \end{pmatrix}$ formed by the three Pauli matrices. For simplicity 
we have assumed here symmetric coupling to the left and right reservoirs.
Besides the quantized spin $\vec{S}$ in the dot, we have introduced fermion
operators describing electrons in the reservoirs, as given by the Hamiltonian:
\begin{equation}
\label{eq:lead_hamiltionian}
H_{0}=\sumk{\sigma\alpha}\ek{\vpht} \cd{\alpha}{} \cv{\alpha}{}, 
\end{equation}
where the operator $\cd{\alpha}{}$ creates one electron in the lead 
$\alpha=L,R$ with momentum $\vec k$, energy $\ek{\vpht}$ and 
spin $\sigma=\uparrow,\downarrow$.  

It is useful to define a ``local'' electronic state of the reservoirs 
$c_{0\sigma}^{\dagger}=\sum_{\vec{k}\alpha} \cd{\alpha}{}/\sqrt{2}$, allowing
to write the Kondo interaction in a more compact way:
\begin{equation}
\label{eq:kondo_alt}
H_{K}= J{\bf S}\;\cdot\!\!
\sum_{\sigma_1\sigma_2}
c_{0\sigma_1}^{\dagger}\frac{\vec{\tau}^{\vpht}_{\sigma_1\sigma_2}}{2}
c_{0\sigma_2}^{\phantom{\dagger}}.
\end{equation}
In this form, a {\it single} screening channel couples to the impurity
spin, which allows to understand the ultimate formation of a non-degenerate
singlet ground state.
The conductance through the dot can be evaluated in the linear response regime 
using Kubo formula~\cite{mahan_many-particle_1990,bruus_many-body_2004} 
\begin{equation}\label{eq:kubo_formula}
\g = \lim_{\omega\rightarrow 0} \frac{1}{\omega}\int_{0}^{\infty}e^{i\omega t}
\left\langle\left[I(t),I(0)\right]\right\rangle dt,
\end{equation}
with the current operator:
\begin{equation}\label{eq:current_operator}
I = \frac{e}{2} \frac{d}{dt}\left(N_R-N_L\right),
\end{equation}
in terms of the charge operator $N_\alpha$ in each lead.
The complete scattering process can be formally described by the full Green's function
of the conduction electrons~\cite{pustilnik_kondo_2004,Glazmann_2001,borda_theory_2007}: 
\begin{eqnarray}
\label{eq:T_matrix}
G^{R}_{c,\sigma}(\vec{k},\vec{k}',\omega)&=&
G^{R}_{0c}(\vec{k},\omega)\delta_{\vec{k},\vec{k}'}\\
\nonumber
&&+G^{R}_{0c}(\vec{k},\omega)\T^{R}_{\sigma}(\omega)
G^{R}_{0c}(\vec{k}',\omega),
\end{eqnarray}
where $\T^{R}_{\sigma}(\omega)$ denotes the retarded $\T$-matrix (which is
momentum independent due to the local nature of the Kondo interaction). We
have also introduced the free Green's function $G^{R}_{0c}(\vec{k},\omega)$ 
of the conduction electrons:
\begin{equation}
G^{R}_{0c}(\vec{k},\omega) = \frac{1}{\omega-\epsilon_{\vec{k}}+i0^+}.
\end{equation}
In all generality, the equilibrium conductance can be related to the
$\T$-matrix by the formula (here in the case of left/right symmetric barriers):
\begin{equation}
\label{eq:T_matrix_time_ordered_imag_maj}
\g= \frac{2e^2}{h} 
 \int_{-\infty}^{\infty} d\ek{}
[-n_F'(\ek{})]
\frac{1}{2}\sum_{\sigma}\left[
-\pi\rho_0\T^{''R}_{\sigma}(\ek{}) 
\right],
\end{equation}
with the density of states at the Fermi level $\rho_0=-(1/\pi) 
\mathcal{I}m\sum_k 1/(-\epsilon_{\vec{k}}+i0^+)$, and the Fermi function $n_F$.
In what follow, the standard notation $\T^{''R}=\mathcal{I}m[T^{R}]$ is used.
Now focusing on the Kondo problem, straightforward derivation using equation
of motion for the conduction electron Green's function~(\ref{eq:T_matrix})
provides the explicit form of the $\T$-matrix:
\begin{eqnarray}
\label{eq:T_matrix_Kondo}
\T_{\sigma}(\tau)&=&-\frac{J}{2}\left\langle S^z \right\rangle  -  \frac{J^2}{4}
\sum_{\sigma_1 \sigma_2}\\
\nonumber
&&\!\!\!\!\times
\left\langle T_{\tau}
\left[c^{\vpht}_{0\sigma_1}(\tau)\vec{\tau}^{\vpht}_{\sigma\sigma_1}\cdot\vec{S}(\tau)
c^{\dagger}_{0\sigma_2}(0)\vec{\tau}^{\vpht}_{\sigma_2\sigma}\cdot\vec{S}(0)
\right]\right\rangle
\end{eqnarray}
using standard imaginary time-ordered correlations functions.
The $\T$-matrix in Eq.~(\ref{eq:T_matrix_Kondo}) describes spin-flip processes to 
all orders, but its perturbative expansion in the small parameter $\rho_0 J$ is
known to be logarithmically divergent, giving rise to the so-called Kondo problem.
The mathematical basis for the systematic resummation of logarithmic terms
in perturbation theory lies in renormalization group ideas, where electron
states of the reservoirs are skimmed progressively from high to low
energy. At leading order, this procedure is encapsulated by the following
flow equation~\cite{hewson_kondo_1997,pustilnik_kondo_2004}:
\begin{equation}
\label{flow}
D \frac{d J}{d D}= \rho_0 J^2,
\end{equation}
where $D$ is the running high frequency cutoff of the conduction band,
starting with the initial value $D_0$ at the band edge (so that 
$\rho_0=1/(2D_0)$ for a purely flat density of states).
Direct integration leads to the renormalized dimensionless Kondo interaction 
at the energy scale $D$:
\begin{equation}
\label{eq:renormj}
\rho_0 J_{\text{R}}=\frac{\rho_0 J}{1-\rho_0 J\log\left|\frac{D_0}{D}\right|}
=\frac{1}{\log\left|\frac{D}{T_K}\right|}
\end{equation}
with the Kondo temperature $T_K=D_0 e^{-\frac{1}{\rho_0 J}}$ which
sets the frontier beyond which perturbation theory breaks down.
In the weak-coupling regime $T\gg T_K$ the conductance $\g(T)$ is given by:
\begin{equation}
\label{eq:conductance_weak}
\g(T)=\frac{2e^2}{h} \frac{3\pi^2}{16} 
\frac{1}{\log^2\left|\frac{T}{T_K}\right|}.
\end{equation} 
The divergence at temperature of the order of $T_K$ marks the formation of
a Kondo resonance, that in reality saturates in the low temperature strong-coupling 
regime towards the unitary limit $\frac{2e^2}{h}$. Capturing the crossover from 
high to low temperature is possible only by non-perturbative methods, such as Wilson's 
numerical renormalization group (NRG)~\cite{wilson_renormalization_1975}.  
The resulting physical picture of the strong-coupling regime is the screening of the local
spin $S=1/2$ by the surrounding conduction electrons~\cite{Coleman_2002}, with
a binding energy for the singlet ground state of the order of $T_K$. Scattering
in the low-temperature regime is then dominated by irrelevant terms, leading to a
Fermi-liquid behavior for $T\ll T_K$:
\begin{equation}
\label{eq:conductance}
\g(T)=\frac{2e^2}{h}\left[1-\frac{\pi^4}{16}\left(\frac{T}{T_K}\right)^2\right].
\end{equation}
Alternatively to the temperature dependence of the equilibrium conductance,
magnetic field and finite bias effect lead similarly to a suppression of
transport when the scale of the perturbation exceeds $T_K$. Focusing on the finite bias
Kondo resonance (at temperatures much below $T_K$), it has been established 
theoretically and experimentally that the Kondo anomaly is robust to a Zeeman
field of the order of $T_K$, which reflects the singlet binding of the ground
state. One goal of the paper is to study the Zeeman splitting of Kondo
resonances associated to a spin-degenerate ground state. A basic model
for the formation of this more exotic state is reviewed now.

\subsection{Spin ${\bm S=1}$ case}

The Kondo Hamiltonian~(\ref{eq:kondo_model}) is straightforwardly generalized 
to higher spin values. In particular, quantum dots with an even number of
electrons may lead to the formation of a spin $S=1$ magnetic state, which
can undergo the so-called underscreeed Kondo 
effect~\cite{nozieres_kondo_1980,Mehta,Koller,Logan,roch_observation_2009,Florens_Review}. 
In the weak coupling regime [See Ref.~\onlinecite{SchurichtAndergassen} for a thorough recent study], 
renormalization of the antiferromagnetic Kondo interaction still follows Eq.~(\ref{flow}), 
and the conductance reads for $T\gg T_K$:
\begin{equation}
\label{eq:conductance_weak_under}
\g(T)=\frac{2e^2}{h} \frac{\pi^2}{8} 
\frac{1}{\log^2\left|\frac{T}{T_K}\right|}.
\end{equation} 
Thus a Kondo resonance develops at $T<T_K$ which can reach up to $2e^2/h$ values 
at low temperature~\cite{Logan}. The crucial difference with respect to
the case $S=1/2$ is that a spin $S=1$ coupled to a single screening channel
cannot bind into a total singlet, so that the resulting ground state has
to accomodate a $\log(2)$ entropy. Nozi\`eres and Blandin~\cite{nozieres_kondo_1980}
showed early on that the scattering process on such a partially screened
state was anomalous. Their argumentation was based to a description of the 
strong coupling fixed point in terms of an effective spin $S^\mathrm{eff}=1/2$
describing the remaining degeneracy in the system. Because the original
spin $S=1$ tends to bind antiparallel spin electron states, Pauli principle
leads to fluctuations with parallel spin electrons, thus leading to
an effective {\it ferromagnetic} interaction $J^\mathrm{eff}<0$, with
$J^\mathrm{eff}\propto - [\rho_0^2 J]^{-1}$ in the mathematical limit
$\rho_0 J\gg1$ considered by Nozi\`eres and Blandin~\cite{nozieres_kondo_1980}.
Integrating the flow equation~(\ref{flow}) with a ferromagnetic coupling leads to 
the renormalized Kondo interaction:
\begin{equation}
\label{eq:renormjFerro}
\rho_0 J^\mathrm{eff}_{\text{R}}=\frac{\rho_0 J^\mathrm{eff}}{1+\rho_0
|J^\mathrm{eff}|\log\left|\frac{D_0}{D}\right|},
\end{equation}
which does not diverge upon decreasing the cutoff $D$, but rather
logarithmically vanishes at low energy. Considering the lowest order contribution
to transport, one then arrives at the following result for the conductance in the 
strong coupling regime $T\ll T_K$:
\begin{equation}
\label{eq:conductance_low_under}
\g(T)=\frac{2e^2}{h}\left[1-\frac{3\pi^2}{16} 
\frac{1}{\log^2\left|\frac{T}{T_K}\right|} \right]
\end{equation}
This slow logarithmic approach to the low temperature fixed point was
observed in recent experiments~\cite{roch_observation_2009,ParksUS}.
Our goal for the rest of the paper is to understand the magnetotransport
of such underscreened Kondo anomalies. Indeed, despite the binding
energy $T_K$, the underscreened ground state is still two-fold degenerate,
and is expected to be highly sensitive to the Zeeman effect (the
spin susceptibility is strictly speaking infinite at zero temperature). 
Before addressing this physical question, we will develop the required methodology.

\section{Majorana diagrammatics for the perturbative Kondo problem}
\label{sec:majo}

\subsection{Majorana representation and observables}

Calculating the perturbative expansion in the Kondo
problem~(\ref{eq:kondo_model}), especially at high orders, requires
the evaluation of correlation functions with multiple occurence
of the spin operators. These involve cumbersome algebra due to the 
peculiar commutation relations of spin operators (which are neither 
fermions nor bosons), which naively precludes the use of the useful Wick's 
theorem. A standard alternative is to represent the quantized spin 
with fictitious particles, for instance using the Abrikosov 
fermion~\cite{mattuck_guide_1992} description of a spin $S=1/2$:
\begin{equation}
\vec{S}=
\sum_{\sigma_1\sigma_2}
f^\dagger_{\sigma_1}
\frac{\vec{\tau}^{\vpht}_{\sigma_1\sigma_2}}{2}
f^{\phantom{\dagger}}_{\sigma_2}
\end{equation}
with a spin-dependent single fermion level $f^\dagger_\sigma$
submitted to the constraint $\sum_\sigma f^\dagger_\sigma
f^{\phantom{\dagger}}_{\sigma}=1$.
While this now allows the use of Wick's theorem, the computation 
of observables is not very much simplified, because a single
spin operator involves the product of two fermions. For instance
the spin correlation functions $\chi_i(\tau)=\langle S^i(\tau)S^i(0)\rangle$
(with $i=x,y,z$) involve four fermions, while the $\T$-matrix 
Eq.~(\ref{eq:T_matrix_Kondo})
requires a correlator with six fermions (including the two electrons originating 
from the leads). Order by order terms can thus be obtained with increasing
effort, but systematic ways to perform resummations of the perturbation
series seems hopeless, especially for the $\T$-matrix, so that the perturbative
renormalization group is often preferred in practice.
Despite the successes of the renormalization ideas, difficulties arise
when considering physical quantities which depend on several energy
scales together (such as temperature, magnetic field, voltage, or frequency),
in which case a single-scale renormalization procedure does not apply,
and more elaborate functional forms of the RG must be
used~\cite{rosch_spectral_2003,rosch_kondo_2005,Kehrein,Schoeller,Schuricht,SchurichtAndergassen}. While these 
method will certainly
perform well for the ferromagnetic Kondo problem that we consider here,
we would like to consider an alternative methodology, which, while not as general
and powerful as the RG, will lead to similarly accurate results by the much simpler 
means of standard Green's function perturbation theory.

The idea that we will follow, building on earlier
works~\cite{parcollet_perturbative_2002,shnirman_spin-spin_2003,mao_spin_2003,bulaevskii_tunneling_2003,Florens_spinboson}
is to use a Majorana fermion representation of a single spin $S=1/2$:
\begin{equation}\label{eq:spin_majorana_definition}
\vec{S} = - \frac{i}{2} \vec{\eta} \times \vec{\eta},
\end{equation}
where $\vec{\eta}=(\eta_1,\eta_2,\eta_3)$ is a triplet of real fermions
which satisfy the anticommutation relation $\{\eta_a,\eta_b\}=\delta_{ab}$.  
One first advantage is for instance that the Hilbert space does not need to be 
restricted, because $\vec{S}^2=\frac{3}{4}$ is automatically 
fulfilled~\cite{parcollet_perturbative_2002,mao_spin_2003,shnirman_spin-spin_2003}. 
Despite the writing of the spin operators in terms of a product of
two Majorana fermions, a real gain can be achieved by the
alternative decomposition~\cite{shnirman_spin-spin_2003,mao_spin_2003}:
\begin{equation}
\vec{S}=\Phi\vec{\eta}~~~~~~\mathrm{with}~~
\Phi= -2i\eta_1\eta_2\eta_3.
\end{equation}
One can then easily check that the fermion $\Phi$ commutes with the
Kondo Hamiltonian, so that for instance the $z$-axis spin correlation function can be 
written:
\begin{equation}
\begin{split}
\chi_z(\tau,0)&=-\left\langle T_{\tau}\left[S^z(\tau)S^z(0)\right]
\right\rangle\\
&= - \left\langle
T_{\tau}\left[\Phi(\tau)\eta^{\vpht}_3(\tau)\Phi^\dagger(0)\eta^{\dagger}_3(0)\right]
\right\rangle \\ &=  G_{\Phi}(\tau)G_{\eta_3}(\tau)
\end{split}
\label{newchiz}
\end{equation}
introducing the Majorana Green's functions:
\begin{equation}\label{eq:Majorana_fermion_propagator_general}
G_{\eta_a\eta_b}(\tau)=-\left\langle T_{\tau}
\left[\eta^{\vpht}_{a}(\tau)\eta^{\dagger}_{b}(0)\right]\right\rangle
\end{equation}
as well as the propagator of the fermion $\Phi$, which remains
{\it free} to all orders in perturbation theory:
\begin{equation}\label{eq:free_fermion}
G_{\Phi}(\tau)=-\langle T_{\tau}
\left[\Phi^{\vpht}_{}(\tau)\Phi^{\dagger}_{}(0)\right]\rangle=-\frac{1}{2}
\mathrm{Sign}(\tau).
\end{equation}
Now the spin susceptibility Eq.~(\ref{newchiz}) is captured just by a
single Majorana propagator, which allows the use of Dyson's equation
instead of more cumbersome response function framework.
Although Majorana fermions are real we distinguish formally $\eta^{\dagger}$ 
and $\eta$. This step makes the diagrammatic expansion clearer once expressed 
in terms of particle and holes~\cite{parcollet_perturbative_2002}. 

Considering now the $\T$-matrix of the Kondo model, originally a six-fermion
correlator, the above arguments leads to a simpler four-fermion 
function:
\begin{eqnarray}
\label{eq:T_matrix_time_ordered_imag_maj_second}
\T_{\sigma}(\tau) &=&- \frac{J^2}{4}G_{\Phi}(\tau)
\sum_{\sigma_1 \sigma_2}\sum_{a b} \\ 
\nonumber
&&\times
\left\langle T_{\tau}
\left[c^{\vpht}_{0\sigma_1}(\tau)\tau^{a}_{\sigma\sigma_1}\eta^{\vpht}_a(\tau)
c^{\dagger}_{0\sigma_2}(0){\tau}^{b}_{\sigma_2\sigma}\eta^{\dagger}_{b}(0)
\right]\right\rangle.
\end{eqnarray}
Up to the constant $\Phi$ propagator, the $\T$-matrix assumes the form
of a response function, for which standard perturbative methodology 
exists, although more involved than for the Majorana propagators required
for the computation of spin dynamics.

\begin{figure}
\includegraphics{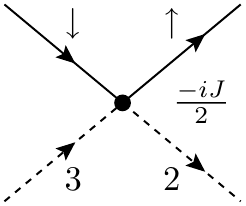}
\caption{ \label{eq:vertex_diagramm}
Interaction vertex associated with the first term in the Kondo
Hamiltonian~(\ref{KondoHamMajo}) expressed in the Majorana language. Full lines 
denote conduction electrons and broken lines Majorana fermions.}
\end{figure}

In order to proceed with perturbation theory, we finally need the Majorana
form of the Kondo Hamiltonian~(\ref{eq:kondo_model}):
\begin{equation}
\label{KondoHamMajo}
\begin{split}
	 H_{K}= &\phantom{+|} \left(\frac{-iJ}{2}\right) \left(
	 c_{0\uparrow}^{\dagger}c^{\vphantom{\dagger}}_{0\downarrow}+c_{0\downarrow}^{\dagger}c^{\vphantom{\dagger}}_{0\uparrow}
	 \right) \eta^{\dagger}_2 \eta^{\vpht}_3\\&+
	 \left(\frac{-\hphantom{i}J}{2}\right) \left(
	 c_{0\uparrow}^{\dagger}c^{\vphantom{\dagger}}_{0\downarrow}-c_{0\downarrow}^{\dagger}c^{\vphantom{\dagger}}_{0\uparrow}
	 \right) \eta^{\dagger}_1\eta_3^{\vpht} \\&+ \left(\frac{-iJ}{2}\right)
	 \left(
	 c_{0\uparrow}^{\dagger}c^{\vphantom{\dagger}}_{0\uparrow}-c_{0\downarrow}^{\dagger}c^{\vphantom{\dagger}}_{0\downarrow}
	 \right)\eta^{\dagger}_1\eta^{\vpht}_2.
\end{split} 
\end{equation}
Interaction vertices between pairs of Majoranas and pairs of conduction
electrons are readily identified from this rewriting of the Hamiltonian,
as shown in Fig.~\ref{eq:vertex_diagramm}.
%

\subsection{Majorana diagrammatics at zero magnetic field}

\begin{figure}
\includegraphics{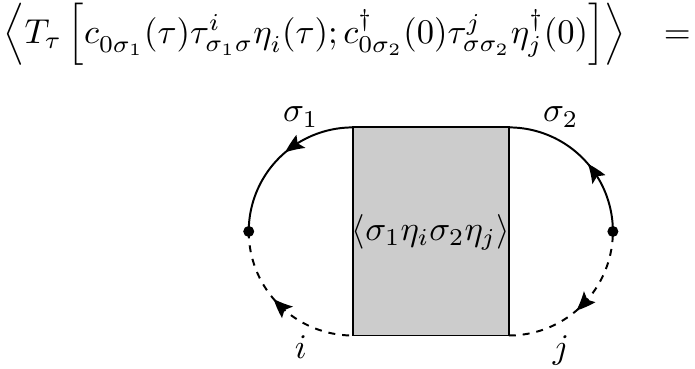}
\caption{ \label{eq:response_diagram} 
Graphical representation of the general correlator involved in the
Majorana decomposition of the $\T$-matrix
Eq.~(\ref{eq:T_matrix_time_ordered_imag_maj_second}) and
Eq.~(\ref{eq:response_matrix}). The shaded box indicates interaction
processes between conduction electrons and Majorana fermions to all
orders in perturbation theory.
}
\end{figure}

In this subsection we will consider the perturbative expansion 
of the $\T$-matrix in terms of Majorana fermions given by
Eq.~(\ref{eq:T_matrix_time_ordered_imag_maj_second}).
The required response functions appearing in the $\T$-matrix are given 
diagrammatically in Fig.~\ref{eq:response_diagram}
where the box represents spin-spin interactions to all orders. 
Since there are three different Majorana fermions, in principle nine different
correlators have to be expanded in perturbation theory, but in
reality only a small number of independent correlation functions arise,
as we will see.
To show this, we define the following abbreviation for the response functions
introduced in Fig.~\ref{eq:response_diagram}:
\begin{equation}\label{eq:response_matrix}
M(\tau)=   
\begin{pmatrix}
\hphantom{-i}			\langle  \downarrow \eta_1 \downarrow	\eta_1\rangle	 	& 
\hphantom{-}i		  	\langle  \downarrow \eta_1 \downarrow 	\eta_2\rangle	  	&
\hphantom{-i}			\langle  \downarrow \eta_1 \uparrow	\eta_3\rangle		\\
-i 				\langle  \downarrow \eta_2 \downarrow 	\eta_1\rangle 	&
\hphantom{-i} 			\langle  \downarrow \eta_2 \downarrow	\eta_2\rangle 	&
-i 				\langle  \downarrow \eta_2 \uparrow	\eta_3\rangle   	\\
\hphantom{-i} 			\langle  \uparrow   \eta_3 \downarrow	\eta_1\rangle		&
\hphantom{-}i			\langle  \uparrow   \eta_3 \downarrow	\eta_2\rangle		&
\hphantom{-i} 			\langle  \uparrow   \eta_3 \uparrow 	\eta_3\rangle
\end{pmatrix},
\end{equation}
which allow to write more compactly the $\T$-matrix:
\begin{equation}
\label{eq:t_matrix_short}
T_{\uparrow}(\tau)=-\frac{J^2}{4}G_{\Phi}(\tau)
\begin{pmatrix}
1, & 1, & 1
\end{pmatrix}\cdot M(\tau) \cdot 
\begin{pmatrix}
1 \\ 1  \\ 1
\end{pmatrix}.
\end{equation}

\begin{figure*}
\centering
\includegraphics{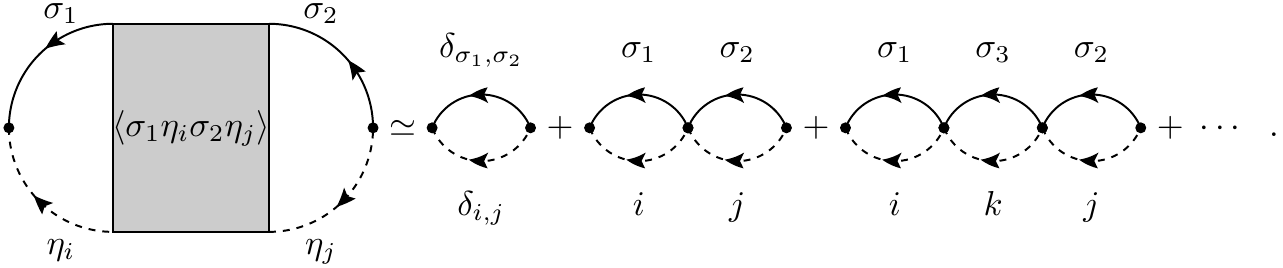}
\caption{\label{fig:t_matrix_response_feynmandiagram_expansion}
RPA-expansion of the response functions entering the $\T$-matrix in the
Majorana language. The single polarization bubble involves the product of
a free Majorana fermion propagator $G_{0\eta}$ and of a
conduction electron propagator $G_{0c}$, and the RPA resummation corresponds to the leading
logarithmic approximation of the standard one-loop renormalization procedure.}
\end{figure*}

In order to perform a leading-log calculation of the $\T$-matrix, we will use
in what follows a RPA resummation with a single polarization bubble, owing
to the fact that non-RPA diagrams ({\it i.e.} involving crossings) are higher 
orders in logarithmic singularities.
In Fig.~\ref{fig:t_matrix_response_feynmandiagram_expansion} this RPA expansion
is shown for the response function entering the $\T$-matrix.
In the case of spin-rotation invariance, all the bubbles in
Fig.~\ref{fig:t_matrix_response_feynmandiagram_expansion} take the
same value (given here in imaginary time):
\begin{equation}
\label{eq:bubble_zero}
\Pi_{0\eta}(\tau)=G_{0c}(\tau) G_{0\eta}(\tau).
\end{equation}
Here the bare Majorana Green's function simply read:
\begin{equation}
G_{0\eta}(i\omega)=\frac{1}{i\omega},
\end{equation}
while the conduction electron propagator are given by:
\begin{equation}
G_{0c}(i\omega)=\sum_{\vec{k}} \frac{1}{i\omega-\epsilon_{\vec{k}}}.
\end{equation}

\begin{figure}
\center
\includegraphics{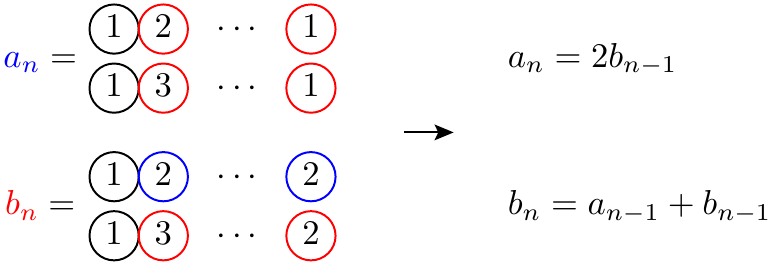}
\caption{\label{fig:recursion-diagram}
(color online) 
Combinatorics of the RPA diagrams at zero magnetic field:  circles denote a single polarization
bubble involving a Majorana fermion with flavor $1$,$2$, or $3$. The coefficient
$a_n$ counts the number of possibilities at order $n$ to end up with
the same index, while the coefficient $b_n$ counts the sequences with different
head and tail. The mathematical recursion on the right side follows from this obvious 
pictorial description.}
\end{figure}

The first terms in the diagrammatic series for the $\T$-matrix are easily 
checked to read:
\begin{equation}\label{eq:result_convolution}
\T_{\uparrow}(\tau)=\frac{J^2}{4}G_{\Phi}(\tau)\left[ 3\Pi_{0\eta} +\frac{J}{2}
6\Pi_{0\eta}*\Pi_{0\eta}+\dots \right](\tau).
\end{equation}
The factors $3$ and $6$ stem from equation~\eqref{eq:t_matrix_short},
because at order $J^2$ the matrix $M$ in Eq.~(\ref{eq:response_matrix}) is 
purely diagonal, while at order $J^3$ it is purely off-diagonal.
The convolution operator in Eq.~(\ref{eq:result_convolution}) is
standardly defined as:
\begin{equation}
\label{eq:bubble_first}
\frac{J}{2}\left[\Pi_{0\eta}*\Pi_{0\eta}\right](\tau)=
\int_0^{\beta}\!\!\!\!\!d\tau_1\; \Pi_{0}(\tau-\tau_1)\Pi_{0\eta}(\tau_1)
\end{equation}
We can already anticipate that a RPA-like resummation will be
possible by Fourier transforming to imaginary frequencies.
However, the numerical coefficient of the series~(\ref{eq:result_convolution})
need to be obtained, and for this we will use a recursion based
on the following diagrammatic rules.
In principle, there are only two different response-functions. The first type
is a response-function where the {\it same} Majorana fermion enters and leaves the 
interacting region (gray box of the $\T$-matrix in 
Fig.~\ref{fig:t_matrix_response_feynmandiagram_expansion}), or equivalently
when no spin-flip of the conduction electrons occurs.
These contributions are the diagonal terms of the matrix given in
equation~\eqref{eq:response_matrix} and we call $a_n$ their combinatorial
factor, so that they contribute in total to $3 a_n$ to the $\T$-matrix in
Eq.~(\ref{eq:t_matrix_short}).
The second type is a response-function where
{\it different} Majorana fermions enter and leave the interacting 
region, so that the conduction electrons have flipped their spin.
These processes are labeled with a combinatorial factor $\textcolor{black}{b_n}$
associated to the non-diagonal terms of the matrix in
Eq.~\eqref{eq:response_matrix}, thus contributing $6b_n$ to the $\T$-matrix,
so that we have the general RPA-series:
\begin{equation}\label{eq:result_convolution_rule}
\T_{\uparrow}(\tau)
=\frac{3J^2}{4}G_{\Phi}(\tau)
\sum_{n=0}^{\infty}(\textcolor{black}{a_n}+2\textcolor{black}{b_n}) \left(\frac J 2
\right)^n\underbrace{\left[\Pi_{0\eta}*\cdots*\Pi_{0\eta}\right]}_{(n+1)\operatorname{times}}
\end{equation}
where $a_0=1,~b_0=0,~a_1=0$ and $b_1=1$. In order to find the explicit 
expression for $a_n$ and $b_n$, we note the combinatorics of Fig.~\ref{fig:recursion-diagram}
which follows from the fact that a Majorana fermion cannot scatter into itself,
due to the identity $\eta_a^2=1/2$. Thus a diagonal process at order $n$
(described by $a_n$) results necessarily from a combination of all non-diagonal
processes at order $n-1$ (described by $b_{n-1}$) and of a single off-diagonal
term, for which two possibilities arise. This shows readily that $a_n=2
b_{n-1}$. Similarly an off-diagonal process at order $n$ can be decomposed either
by an arbitrary diagonal processes at order $n-1$ and a single off-diagonal
one, or by all possible non-diagonal processes at order $n-1$ followed by a single
non-diagonal one, so that $b_n=a_{n-1}+b_{n-1}$. We can trivially solve this
recursion, and we find $a_n+2b_n=2^n$.
Using this result, the $\T$-matrix is given at RPA level by
\begin{equation}\label{eq:result_convolution_rule_final}
\T_{\uparrow}(\tau)= \frac{3J^2}{4}G_{\Phi}(\tau)\Pi_{\eta}(\tau), 
\end{equation}
where $\Pi_{\eta}(\tau)$ is simply determined by a geometric series in Matsubara
frequencies:
\begin{equation}
\label{eq:sum_omega}
\Pi_{\eta}(i\nu)= \frac{\Pi_{0\eta}(i\nu)}{1-J\Pi_{0\eta}(i\nu)}.
\end{equation}

\begin{figure}
\centering
$\vpht$\\[0.2cm]
\includegraphics[width=0.47\textwidth]{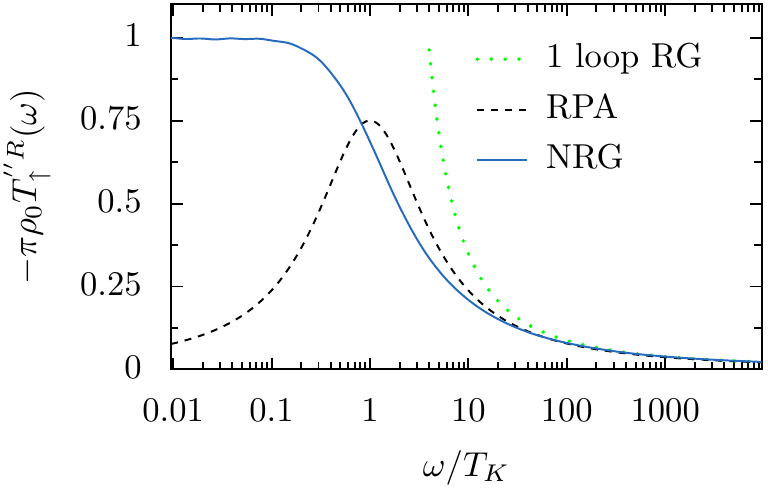}
\caption{\label{fig:T_matrix_RPA_withoutb}
(color online) $\T$-matrix of the Kondo model with antiferromagnetic exchange at
zero magnetic field, comparing the NRG simulations (full line) to the poor man's
scaling one-loop result (dotted line) and to the Majorana diagrammatics
at RPA level (dashed line) given by Eq.~(\ref{eq:T_mat_final}).}
\end{figure}

We finally compute the expression for the $\T$-matrix, first by
Fourier transforming to fermionic Matsubara frequencies:
\begin{equation}
\label{eq:T_fouriertansform_1}
\begin{split}
\T_{\uparrow}(i\omega_n) &=
 \frac{3J^2}{4}\frac{1}{\beta}\sum_{i\nu_1}G_{\Phi}(i\nu_1)\Pi(i\omega_n-i\nu_1)\\
			&=
 \frac{3J^2}{8\pi}\int_{-\infty}^{\infty}\!\!\!\! d\omega_2\;
 \Pi^{''R}_{\eta}(\omega_2)\frac{1+2n_B(-\omega_2)}{\omega_2-i\omega_n},
\end{split}
\end{equation}
and using a spectral decomposition ($n_B$ is the Bose distribution).
The analytic continuation is then straightforward, and we obtain the imaginary
part of the real-frequency $\T$-matrix:
\begin{equation}\label{eq:T_imag}
\T^{''R}_{\uparrow}(\omega)
	=-\frac{3J^2}{8}\coth\left(\frac{\beta\omega}{2}\right)
	\operatorname{Im}\left(\frac{\Pi^R_{0\eta}(\omega)}{1+J\Pi^R_{0\eta}(\omega)}\right)
\end{equation}
with $\beta=1/T$ the inverse temperature.
The polarization diagram $\Pi_{0\eta}$ is computed by similar
means from Eq.~(\ref{eq:bubble_zero}) and reads for a flat
conduction band with density of states $\rho_0=1/(2D)$ and half-bandwidth $D$:
\begin{equation}\label{eq:polarization_diagram_derivation}
 \Pi_{0\eta}^{R}(\omega) =
\frac{\rho_0}{2}\log\left(\left|\frac{\omega^2-D^2}{\omega^2}\right|\right)
+i\pi\frac{\rho_0}{2}\operatorname{Sign}(\omega)\Theta(\omega^2-D^2)
\end{equation}
leading to the final result for $\T$-matrix at $T=0$:
\begin{equation}\label{eq:T_mat_final}
-\pi\rho_0 \T^{''R}_{\uparrow}(\omega)=
\frac{3\pi^2}{16}\frac{(\rho_0 J)^2 \Theta(\omega^2-D^2)}
{\left(1-\frac{\rho_0J}{2}
\log\left|\frac{D^2-\omega^2}{\omega^2}\right|\right)^2+\frac{\pi^2 (\rho_0 J)^2}{4}}.
\end{equation}
Inserting this expression into the conductance
formula~(\ref{eq:T_matrix_time_ordered_imag_maj}), we recover the
standard high-temperature result Eq.~(\ref{eq:conductance_weak}) for
$T\gg T_K$. Let us contrast our expression to the standard weak-coupling
renormalization group result in the case of antiferromagnetic Kondo
exchange: $ -\pi\rho_0 \T^{''R}_{\uparrow}(\omega)=
\frac{3\pi^2}{16}[\log(\omega/T_K)]^{-2}$. In the range $T_K\ll \omega\ll D$,
both are completely equivalent. For frequencies near the high energy cutoff 
$D$, the correct band-edge contribution appears in Eq.~(\ref{eq:T_mat_final}),
which is usually neglected in the RG flow (but can be in principle incorporated).
Stronger differences occur in the regime near and below $T_K$, where a maximum occurs
in Eq.~(\ref{eq:T_mat_final}) instead of the usual divergence found in one-loop RG. 
The reason is the presence of a constant offset of order $J^2$ in the denominator, 
coming from the imaginary part of the Majorana bubble, which cuts off the Kondo
singularity. This term shows that our RPA scheme picks not only the leading
logarithms, but also some non-logarithmic terms in the perturbation series.
The strong differences between one-loop RG and RPA at frequencies lower than
$T_K$ appear however only in a regime
where both theories are uncontrolled, and is thus of no physical significance.
Indeed, we emphasize that perturbation theory still breaks down in our scheme
for $\omega\ll T_K$, see Eq.~(\ref{eq:T_mat_final}), because the effective dimensionless 
Kondo coupling has reached values of order 1.
Yet, the convergence of our result to NRG data for $\omega\simeq T_K$ seems slightly 
better than the lowest order RG expression, see Fig.~\ref{fig:T_matrix_RPA_withoutb}, 
thanks to this partial resummation of subleading logarithmic terms.
We now show how the Majorana diagrammatics can be extended to include
the Zeeman effect as well.

\subsection{Majorana diagrammatics at finite magnetic field}

\begin{figure}
\includegraphics[scale=0.68]{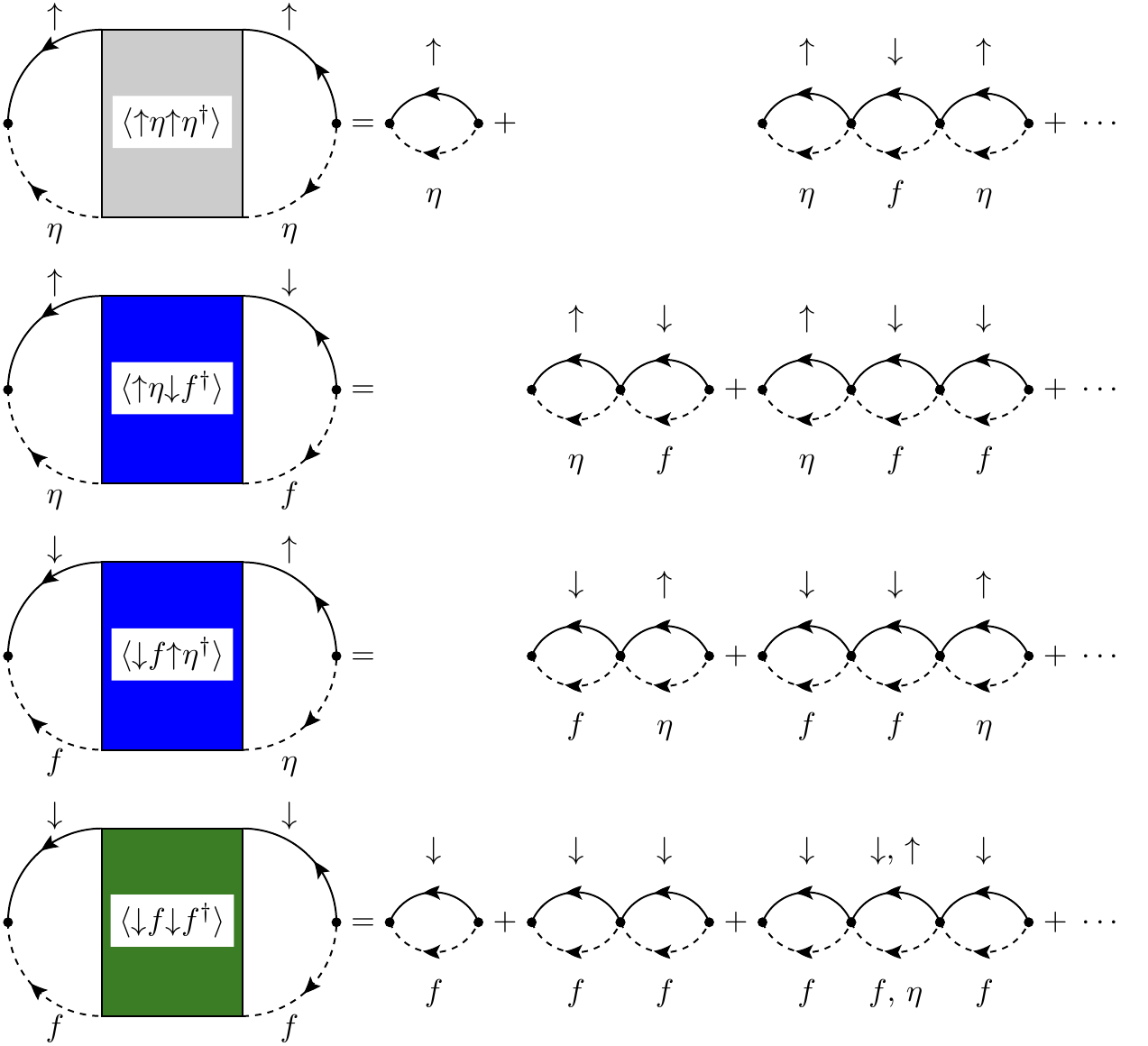}
\caption{\label{fig:RPA_Zeeman} 
(color online) Lowest order contributions to the RPA
evaluation of the $\T$-matrix in the case of finite magnetic field.}
\end{figure}

The Zeeman effect can be easily studied in quantum dots, by applying a 
magnetic field in the plane in the case of semiconducting 
two-dimensional electron gas devices. We note that the latter systems have
relatively small $g$-factor, but because of the low Kondo temperature
that they can achieve (in the range 100mK to 1K), a competition with the 
Kondo effect can occur at magnetic fields of a few Tesla. In molecular quantum
dots larger $g\simeq2$ values are obtained, which helps realizing 
the competition between Kondo and Zeeman effects.
In this section, we simply add the Zeeman term to the Kondo
Hamiltonian~(\ref{eq:kondo_model})
\begin{equation}\label{eq:zeeman_term}
H_{Z}=BS^{z}=-iB\eta_1\eta_2,
\end{equation}
where $B$ is measured in units of the Zeeman splitting. 
We note that the Majorana fermion propagator $G_{ab}(\tau)$ has now non-diagonal 
elements, which makes it more difficult to carry out the previous RPA for the $\T$-matrix.
The first idea is to transform the Majorana fermions so that their propagator becomes
diagonal, which can be done by introducing one real fermion:
\begin{equation}\label{eq:Majorana_f}
f^{\dagger}=\frac{1}{\sqrt{2}}(\eta_1+i\eta_2),~~~
\left\{f^{\dagger},f\right\}=1.
\end{equation}
The Kondo Hamiltonian is given after this transformation by:
\begin{equation}\label{eq:Kondo_trafo}
\begin{split}
H_K=&-\frac{J}{2}\sqrt{2}\left(
c^{\dagger}_{0\downarrow}c^{\vphantom{\dagger}}_{0\uparrow}f^{\dagger}\eta^{\vpht}_3
+c^{\dagger}_{0\uparrow}c^{\vphantom{\dagger}}_{0\downarrow}\eta_3^{\dagger}f
\right) \\ &+
\frac{J}{2}    
\left(
c^{\dagger}_{0\uparrow}f^\dagger fc^{\vphantom{\dagger}}_{0\uparrow}
-c^{\dagger}_{0\downarrow}f^\dagger fc^{\vphantom{\dagger}}_{0\downarrow}
\right)\\
&- \frac{J}{4}[c^{\dagger }_{0\uparrow} c^{\phantom{\dagger}}_{0\sigma\uparrow }
-c^{\dagger}_{0\downarrow} c^{\phantom{\dagger}}_{0\sigma\downarrow }].
\end{split}
\end{equation}
The last term, corresponding to a small magnetic field acting on
the bath of conduction electrons, is expected to remain perturbatively
small. However, we will see in Sec.~\ref{sec:spin} that it needs to be kept in 
order to describe properly the self-energy in second order. 
The Zeeman term now becomes purely diagonal:
\begin{equation}\label{eq:zeeman_trafo}
H_z=Bf^{\dagger}f-\frac{1}{2}B,
\end{equation}
where the last term is only a constant and will be dropped from now on. 
Thus the transformation leads to two different fermionic propagators 
\begin{equation}\label{eq:Majorana_fermion_propag_trafo}
G_{f}(i\omega_n)=\frac{1}{i\omega_n-B},
~~~G_{\eta_3}(i\omega_n)=\frac{1}{i\omega_n},
\end{equation}
where $G_f$ is a canonical fermion propagator and $G_{\eta_3}$ is the propagator
of the unchanged Majorana fermion $\eta_3$. The label $3$ can be dropped and we
will use in what follows $G_{\eta}$ as the Majorana fermion propagator.  
Next, the $\T$-matrix is expressed in terms $f$ and $\eta$:
\begin{equation}\label{eq:trafo_T}
\begin{split}
\T_{\uparrow}(\tau)=&-\frac{J^2}{4}G_{\Phi}(\tau)\left(
2\textcolor{black}{\langle \downarrow f \downarrow f^{\dagger} \rangle} 
+ \sqrt{2}\textcolor{black}{\langle \downarrow f \uparrow \eta \rangle}\right.
\\ 
&+ \left. \sqrt{2}\textcolor{black}{\ \langle \uparrow \eta \downarrow f^{\dagger}
\rangle}
+ \langle \uparrow \eta \uparrow \eta \rangle
\right),
\end{split}
\end{equation}
We have now two different bare polarization bubbles  $\Pi_{0f}$ and 
$\Pi_{0\eta}$, associated to $f$ and $\eta$ respectively, which are 
readily evaluated:
\begin{eqnarray}
\label{polarizationZeemaneta}
\Pi_{0\eta}(\omega)&=&  \frac{\rho_0}{2}
\log\left|\frac{D^2-\omega^2}{\omega^2}\right| + i \frac{1}{2}\pi \rho_0
\operatorname{Sign}(\omega) \\
\label{polarizationZeemanf}
\Pi_{0f}(\omega)&=& \rho_0  \log \left|\frac{D+B-\omega}{B-\omega}\right| +
i \pi \rho_0 \theta(\omega-B),
\end{eqnarray}
in the case $B>0$. Because of these two different contributions, the
combinatorics cannot be guessed as easily as in the previous section,
so we must use a more general resummation technique. The general structure
of perturbation theory (with  bare polarization bubbles) can be considered
at the light of Fig.~\ref{fig:RPA_Zeeman}.
We introduce a dimensionful coefficient $A_n$ which encodes processes at
order $n$ that start and finish with the bubble $\Pi_{0f}$, and a coefficient
$B_n$ that involve $\Pi_{0f}$ and $\Pi_{0\eta}$ at its extremities.
Note that these coefficients are not only combinatoric in character (like
the coefficients $a_n$ and $b_n$ defined in the previous section),
but also include a sum of terms of the type $[\Pi_{0f}]^p [\Pi_{0\eta}]^{n-p}$
with integer $p$, due to the fact that we have two independent polarization
bubbles now. The correlations appearing in Eq.~(\ref{eq:trafo_T}) are then
expressed as:
\begin{equation}\label{eq:last_response}
\begin{split}
&\langle\uparrow\eta\uparrow\eta\rangle 
=\Pi_{0\eta} + 2 \left(\frac{J\Pi_{0\eta}}{2}\right)^2
\sum_{n=0}^{\infty}A_n\left(\frac{J}{2}\right)^{n}\\
&2\langle \downarrow f \downarrow f^{\dagger}
\rangle=2\sum_{n=0}^{\infty}A_n\left(\frac{J}{2}\right)^n\\
&\sqrt{2}\left( \langle \uparrow \eta \downarrow
f^{\dagger}\rangle+ \langle \downarrow f \uparrow \eta
\rangle\right)=2\sqrt{2}\sum_{n=0}^{\infty}B_n\left(\frac{J}{2}\right)^n
\end{split}
\end{equation}
so that the complete RPA series~(\ref{eq:trafo_T}) for the $\T$-matrix
reads:
\begin{equation}\label{eq:T_imagZeeman}
\begin{split}
T^{''R}_{\uparrow}(\omega)
&=-\frac{J^2}{8}\coth\left(\frac{\beta\omega}{2}\right)\Pi^{''R}(\omega)\\
\Pi^R(\omega) =& \Pi_{0\eta}^R(\omega) + 2
\sum_{n=0}^{\infty}\left(\frac{J}{2}\right)^n
\Bigg\{A_n+\left(\frac{J\Pi_{0\eta}^R(\omega)}{2}\right)^2 A_n\\
& + \sqrt{2}B_n\Bigg\}.
\end{split}
\end{equation}

\begin{figure}
\center
\includegraphics{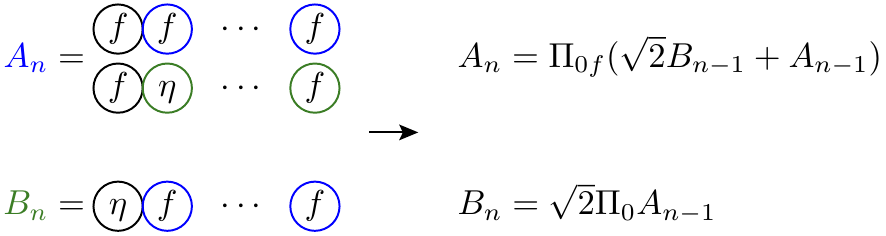}
\caption{\label{fig:recursion_diagram_f_eta}
(color online) 
Combinatorics of the RPA diagrams at finite magnetic field: circles 
denote a single polarization bubble involving a real fermion $f$ or a Majorana
$\eta$. The coefficient $A_n$ picks the contributions at order $n$ that end up with
the same index $f$, while the coefficient $B_n$ collects the sequences with different
head and tail. The mathematical recursion on the right side follows from this
obvious pictorial description.}
\end{figure}

A recursion is easily established on the basis of
Fig.~\ref{fig:recursion_diagram_f_eta}, which can be best cast and 
solved in matrix form:
\begin{equation}\label{eq:recursion_relation_f_eta}
\begin{split}
\begin{pmatrix}
A_n \\ B_n
\end{pmatrix}
&=
\begin{pmatrix}
\Pi_{0f} & \sqrt{2}\Pi_{0f} \\
\sqrt{2}\Pi_{0\eta} & 0 \\
\end{pmatrix}
\cdot
\begin{pmatrix}
A_{n-1} \\ B_{n-1}
\end{pmatrix} \\
&=
\begin{pmatrix}
\Pi_{0f} & \sqrt{2}\Pi_{0f} \\
\sqrt{2}\Pi_{0\eta} & 0 \\
\end{pmatrix}^n\cdot
\begin{pmatrix}
A_{0} \\ B_{0}
\end{pmatrix} \\
\end{split}
\end{equation}
with $A_0=\Pi_{0f}$ and $B_0=0$ as can be inferred from the lowest
order perturbative expansion shown in Fig.~\ref{fig:RPA_Zeeman},
Inserting Eq.~(\ref{eq:recursion_relation_f_eta}) into the
expression~(\ref{eq:T_imagZeeman}) for the $\T$-matrix, we end up with
the final compact result at zero temperature:
\begin{equation}
\label{eqTmag}
T^{''R}_{\uparrow}(\omega) = - \frac{J^2}{8} \operatorname{Sign}(\omega)
\operatorname{Im}\left(\frac{\Pi_{0\eta}+2\Pi_{0f}+\frac{3}{2}J\Pi_{0f}\Pi_{0\eta}}
{1-\frac{J}{2}\Pi_{0f}-\frac{J^2}{2}\Pi_{0\eta}\Pi_{0f}}\right)
\end{equation}
which will be investigated in Sec.~\ref{sec:results}.
If we take the limit $B\to0$ {\it before} the limit of zero temperature,
we find $\Pi_{0\eta}=\Pi_{0f}$, so that expression~(\ref{eqTmag}) 
reduces to
\begin{equation}
\begin{split}
T^{''R}_{\uparrow}(\omega)
&=-\frac{J^2}{8} \operatorname{Sign}(\omega)
\operatorname{Im}
\left(
\frac{3\Pi_{0\eta}\left(1+\frac{J}{2}\Pi_{0\eta}\right)}
{\left(1-J\Pi_{0\eta}\right)\left(1+\frac{J}{2}\Pi_{0\eta}\right)}
\right)
\\      &=
-\frac{J^2}{8} \operatorname{Sign}(\omega)
\operatorname{Im}
\left(
\frac{3\Pi_{0\eta}}{1-{J}\Pi_{0\eta}}
\right)
\end{split}
\end{equation}
which is exactly the one established previously in Eq.~(\ref{eq:T_imag})
for zero magnetic field.
Note however that the limit $B\to0$ and $T\to0$ clearly do not commute
by comparing Eq.~(\ref{polarizationZeemaneta}-\ref{polarizationZeemanf}) and
Eq.~(\ref{eq:polarization_diagram_derivation}), due to the abrupt change in
the ground state degeneracy of the free impurity spin as soon as a small 
magnetic field is turned on. At zero magnetic field, one can check that the expansions
using a spin rotation-invariant formalism or a symmetry-broken one
provide equivalent results for the physical observables (at $B=0$), such as the total
$\T$-matrix $[\T^{R}_{\uparrow}+ \T^{R}_{\downarrow}]/2$.

\section{Results}
\label{sec:results}

In this section we investigate the analytic expressions obtained previously,
both for antiferromagnetic and ferromagnetic exchange, and compare them
to numerical renormalization group (NRG)
calculations~\cite{wilson_renormalization_1975,krishna-murthy_renormalization-group_1980,
bulla_numerical_2008}. The simulations will be performed with 
the full density matrix using the DM-NRG algorithm~\cite{hofstetter_generalized_2000,toth_density_2008}, 
which proved more accurate to compute observables in a finite (and possibly large) 
magnetic field compared to the standard NRG procedure.

\subsection{Anti-ferromagnetic Kondo model}
\label{sec:anti}

The weak coupling RPA result Eq.~(\ref{eq:T_mat_final}) for the $\T$-matrix obtained 
at zero magnetic field can be rewritten in the limit
$D \gg |\omega| \gg T_K$ in terms of the Kondo temperature $T_K$:
\begin{equation}\label{eq:T_mat_final_weak}
-\pi\rho_0\T^{''R}_{\uparrow}(\omega)\approx\frac{3\pi^2}{16}\frac{1}{\log^2\left|\frac{\omega}{T_K}\right|+\frac{\pi^2}{4}}.
\end{equation}
This recovers the standard poor man's scaling result, with some inclusion
of high logarithmic corrections, which cut-off the divergence at $\omega=T_K$
(nevertheless, the RPA breaks down in this regime as well).
This difference between the two results stems from the fact that the imaginary 
part of the polarization diagram is not neglected for the renormalization of 
Kondo exchange in the RPA. A benchmarking of this expression to the DM-NRG calculation 
was already provided in Fig.~\ref{fig:T_matrix_RPA_withoutb}.

\begin{figure}
\includegraphics{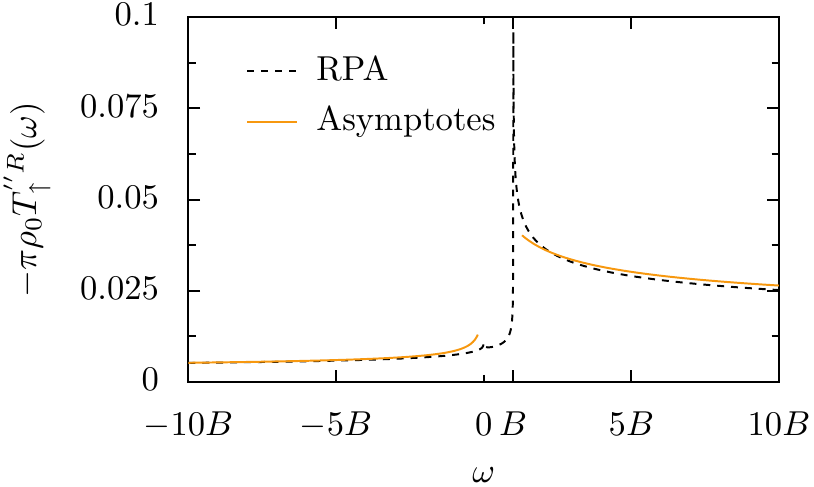}
\caption{\label{fig:result_mag}(color online) Spin-resolved $\T$-matrix of
the Kondo model with antiferromagnetic exchange in a finite magnetic field 
$B=5000 T_K$ as computed in the Majorana diagrammatics at RPA level
from Eq.~(\ref{eqTmag}). The high energy asymptotes given in Eq.~(\ref{eq:asymptotes})
are also shown. A remanence of the Kondo peak occurs for frequencies near $\omega=B$, see 
details in Fig.~\ref{fig:result_1}.}
\end{figure}

We now concentrate on the case of a finite magnetic field, where the presence
of two scales, $\omega$ and $B$, prevent a simple poor man's scaling, and
requires more sophisticated analytical renormalization group
techniques~\cite{rosch_spectral_2003,rosch_kondo_2005}.
Our result Eq.~(\ref{eqTmag}) obtained with the Majorana can be re-expressed
in the regime $D \gg \omega \gg B$ or  $-D \ll  \omega \ll B$ as: 
\begin{equation}\label{eq:asymptotes}
\begin{split}
-\pi \rho_0\T^{''R}_{\uparrow}(\omega) \approx & 
\pi^2 \frac{5}{16 \log^2\left(\left|\frac \omega {T_K} \right|\right)}
~~~~\mathrm{for}\;\hphantom{-} D \gg \omega \gg B
\\ -\pi\rho_0\T^{''R}_{\uparrow}(\omega) \approx &  \pi^2 
\frac{1}{16 \log^2\left(\left|\frac \omega {T_K} \right|\right)}
~~~~\mathrm{for}\; - D \ll \omega \ll B
\end{split}
\end{equation}
resulting in different asymptotes in the case of large positive and negative
frequencies, in agreement with the fRG results by 
Rosch {\it et al.} \cite{rosch_spectral_2003,rosch_kondo_2005}. 
A global view on the the $\T$-matrix is given in Fig.~\ref{fig:result_mag},
which shows a resonance that is now centered at $\omega=B$.
Note that perturbation theory is adequate because the scattering process
remains far away from the unitary limit, $-\pi\rho_0\T^{''R}(\omega)\ll 1$.

\begin{figure}
\includegraphics[width=0.44\textwidth]{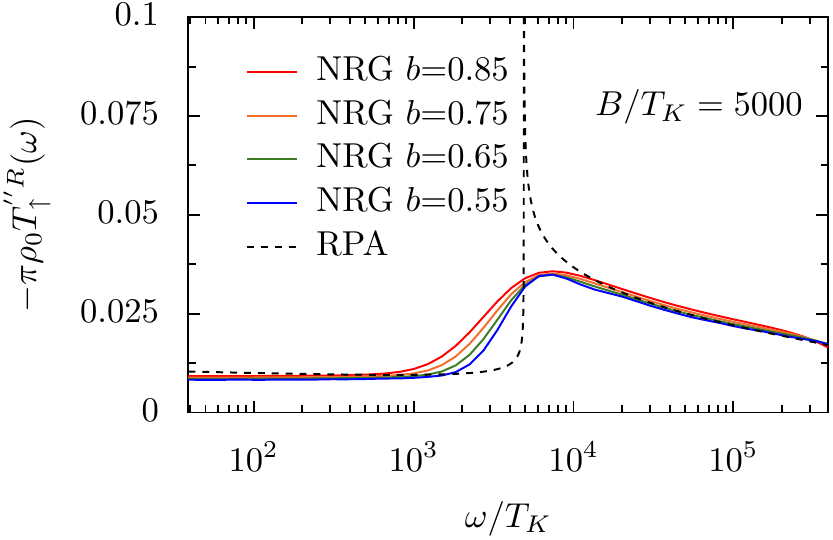}
\caption{\label{fig:result_1} (color online) Close up on the finite magnetic field resonance in
the Kondo model with antiferromagnetic exchange. This is a comparison between 
Majorana diagrammatics at RPA level (dashed line) and DM-NRG simulations with 
decreasing broadening parameter $b=0.85,0.75,0.65,0.55$ (solid lines, top to 
bottom) for a ratio $B/T_K=5000$. The NRG data are clearly overbroadened for 
frequencies near the threshold $\omega=B$. One note also a slight overshoot of the
Majorana result at low frequency, also visible in Fig.~\ref{fig:result_mag}, an
artifact which is discussed in the text. In all other frequency ranges, the
agreement between analytics and numerics is quantitative.}
\end{figure}

Let us now consider the analytical results in more detail, by comparing them to
DM-NRG simulations. For all the NRG calculations, we take the half bandwidth
$D=1$ and the Kondo coupling $J/D=0.145$, so that the Kondo temperature
is $T_K/D\simeq 10^{-6}$. All results will be obtained using the $U(1)$ symmetry
and keeping the first 320 states, using  a logarithmic discretization of the
band with NRG parameter $\Lambda=2$. We consider also the standard 
broadening parameter $b$ in the range 0.5 to 0.9, in order to check possible
problems of convergence in the broadening method (which can arise especially near 
the resonance at $\omega=B$). Fig.~\ref{fig:result_1} shows the comparison of
the NRG calculation and the analytical expression Eq.~(\ref{eqTmag}) for 
the $\T$-matrix.
Clearly the agreement is overall excellent, but let us discuss rather the
nature of the visible discrepancies. The most evident disagreement occurs near
the resonance at $\omega=B$, where the NRG shows a very broad shoulder
while the Majorana result at RPA level presents a threshold followed by a 
spurious logarithmic divergence.
The study of several broadening parameters in Fig.~\ref{fig:result_1}
shows that the broadening of the NRG data converges well except near
the resonance, so that the actual result is likely to display a threshold-like
feature at $\omega=B$ similar to the Majorana result. This broadening issue of
the NRG is a notoriously difficult problem, to which more 
refined broadening methods~\cite{oliveira_generalized_1994,freyn_optimal_2009} do not bring much
improvement~\cite{Florens_spinboson,Weichselbaum_Bfield}. In contrast to
this dynamically generated resonance of the $\T$-matrix, atomic like resonances
(which occur in the abssence of dissipative coupling to the electronic
environment) can be tackled by improved broadening
method~\cite{freyn_optimal_2009,Florens_spinboson,Florens_Review}, as we will see
in Sec.~\ref{sec:spin} for the transverse spin susceptibility. Despite this flaw of the 
NRG to predict a threshold at $\omega=B$, the NRG simulation
predict correctly a finite amplitude peak above the threshold, while the
Majorana diagrammatics leads to a spurious logarithmic divergence at $\omega\to B^+$,
which can be clearly traced back to the form of the $f$-fermion bubble in
Eq.~(\ref{polarizationZeemanf}). It has been argued in the context of
fRG~\cite{rosch_spectral_2003,rosch_kondo_2005} that this spurious singularity 
should be smoothened by taking into account spin relaxation effects, an issue that 
we examine in Sec.~\ref{sec:spin}.

\begin{figure}
\includegraphics[width=0.45\textwidth]{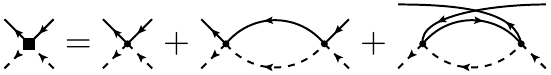}
\includegraphics{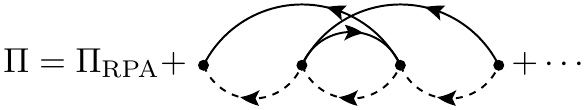}
\caption{\label{fig:one_loop}
Upper panel: standard one-loop renormalization of the Kondo exchange.
Resummation of the top right diagram (particle-hole channel) can be generalized in
the Majorana diagrammatics in terms of the non-RPA extra contribution shown in the lower
panel. The neglect of such terms which remain logarithmic in the range $\omega\ll
T_K$ for finite magnetic field is the likely origin for the discrepancies of the RPA
at low frequency (and finite $B$) with respect to one-loop fRG calculations.}
\end{figure}

A second drawback of the present RPA resummation scheme can be observed in
the low frequency range in Figs.~\ref{fig:result_mag} and \ref{fig:result_1}.
While the plateau of the $\T$-matrix in the range $T_K\ll \omega \ll B$ is
quantitatively reproduced by the Majorana diagrammatics, a spurious logarithmic 
divergence occurs at $\omega\ll T_K$, which leads to progressive deviations
from the numerical results at lowering frequency. The origin of this behavior
is easily understood from the Majorana fermion bubble
Eq.~(\ref{polarizationZeemaneta}), which displays an unbounded logarithm in
the zero frequency limit. We have checked however that including spin relaxation 
effects will not cure this divergence, which has rather to do with deficiencies
of the RPA resummation. Indeed, while the RPA diagrams reproduce exactly the one-loop 
poor man's scaling approach at zero magnetic field, the RPA scheme at finite magnetic 
field is actually not fully equivalent to the fRG approach in the limit $\omega \ll T_K$ 
(the two approaches give however equivalent results otherwise).
This difference can be understood from the fact that a one-loop renormalization 
of the Kondo exchange takes into accounts non-RPA diagram, such as seen in
Fig.~\ref{fig:one_loop}.
One can check that these non-RPA diagram remain logarithmic in the range
$\omega\ll T_K$ for finite magnetic field, and these will likely cancel out
the spurious low frequency divergence of the RPA scheme (these diagrams are
however logarithmically subdominant at {\it high} energy, allowing the good agreement 
between RPA and fRG for the other frequency regimes). We have unfortunately not been
able to resum exactly to all orders the non-RPA diagram shown in the lower
panel of Fig.~\ref{fig:one_loop}. However, it is possible to put the present 
argumentation on a more precise ground, by noting that the leading contribution 
of the shown diagram in Fig.~\ref{fig:one_loop} is given by the combination 
$[\Pi_{0\eta}(\omega)]^2\Pi_{0f}(\omega=0)$. Iterating these terms in a geometric
series to all orders amount to cancel the remaining spurious logarithmic
divergence at low energy, as given by this new form of the $\T$-matrix:
\begin{eqnarray}
\label{eqTnonRPA}
T^{''R}_{\uparrow}(\omega) &=& - \frac{J^2}{8} \operatorname{Sign}(\omega) \times \\
\nonumber
&& \operatorname{Im}\left(\frac{\Pi_{0\eta}+2\Pi_{0f}+\frac{3}{2}J\Pi_{0f}\Pi_{0\eta}}
{1-\frac{J}{2}\Pi_{0f}-\frac{J^2}{2}[\Pi_{0\eta}\Pi_{0f}-\Pi^{'}_{0\eta}\Pi_{0f}(0)]}\right).
\end{eqnarray}
\begin{figure}
\includegraphics[width=0.40\textwidth]{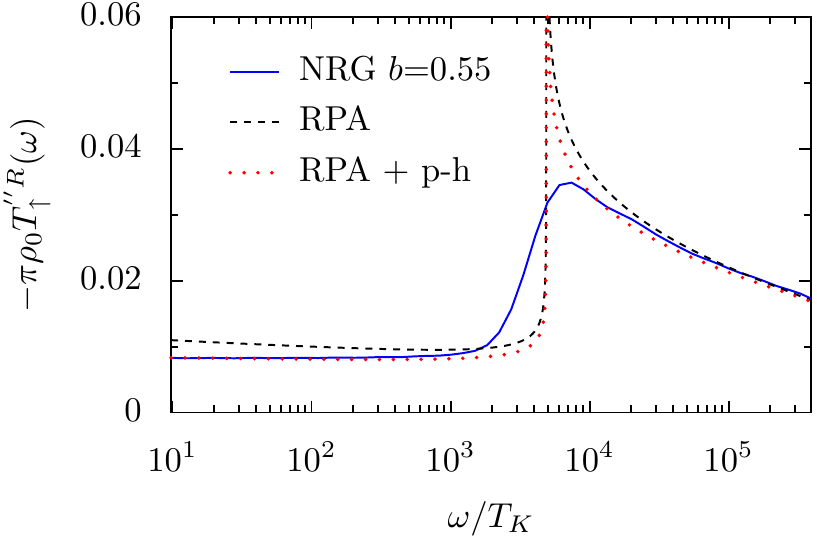}
\caption{\label{fig:RPA_corrected} (color online) Kondo resonance in a finite magnetic field
for antiferromagnetic exchange, with the same parameters as
Fig.~\ref{fig:result_1}. This plot shows the corrected
RPA with the approximate particle-hole channel (RPA+p-h) diagram of 
Eq.~(\ref{eqTnonRPA}), with comparison to NRG calculations and the uncorrected RPA.}
\end{figure}
Figure~\ref{fig:RPA_corrected} shows indeed the elimination of the spurious
low-$\omega$ resonance, and quantitative agreement with the NRG for all
frequencies (except right at the threshold $\omega=B$).
This successful benchmarking of the Majorana diagrammatics allows us to
consider now with confidence the $\T$-matrix in the case of ferromagnetic
exchange.

\subsection{Ferromagnetic Kondo model}\label{sec:ferro}

We now turn to the central question of the present work, namely the
magnetoconductance of the spin $S=1/2$ Kondo model with ferromagnetic exchange.
This description also applies to the strong coupling limit of the spin $S=1$ underscreened
Kondo model with antiferromagnetic
exchange~\cite{nozieres_kondo_1980,cragg_kondo_1978}, recently observed in 
molecular quantum dots~\cite{roch_quantum_2008,roch_observation_2009,ParksUS}.
For both models, the ground state remains a spin doublet with logarithmic
decoupling from the bath of conduction electrons, and so one expects an
extreme sensitivity to an applied magnetic field.
\begin{figure}
\center
\includegraphics[width=0.40\textwidth]{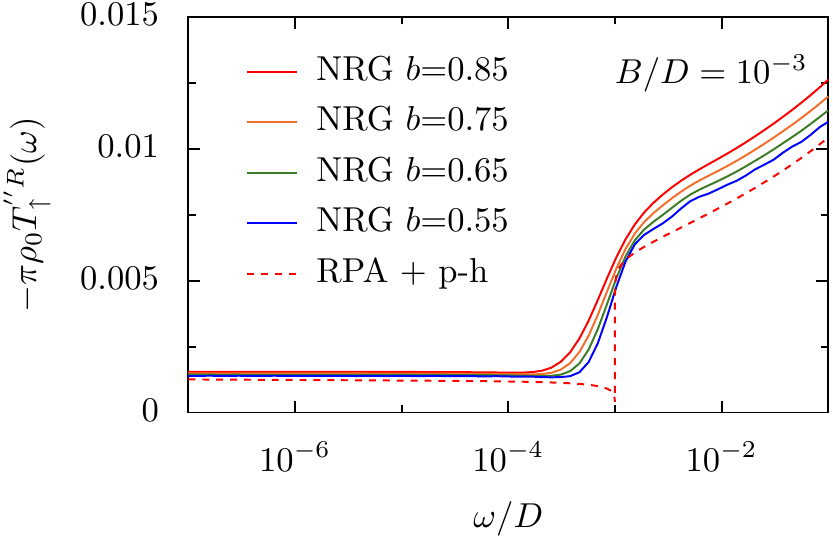}
\caption{\label{fig:ferro1} (color online)
$\T$-matrix of the Kondo model with ferromagnetic exchange $J/D=-0.138$ and finite
magnetic field $B/D=0.001$, comparing the DM-NRG calculations for broadening 
parameters $b=0.85,0.75,0.65,0.55$ (top to bottom) to the RPA result with 
particle-hole corrections Eq.~(\ref{eqTnonRPA}).}
\end{figure}
The ferromagnetic exchange modifies dramatically the flow equation 
of the Kondo coupling, which remains small at all energies, and even
renormalizes to zero in the low frequency limit ~\cite{cragg_kondo_1978,nozieres_kondo_1980}.
We compare again the RPA results (with particle-hole corrections) with DM-NRG
simulations, see Fig.~\ref{fig:ferro1}. We find excellent agreement between the numerics 
and Majorana perturbation theory, as expected from the perturbative nature of the ferromagnetic 
Kondo model. 
However, near the resonance at $\omega=B$, one observes again that
the DM-NRG results are overbroadened, similar to the antiferromagnetic case,
so that the step predicted by the theory is difficult to reproduce.
Perturbation theory displays also an artifact, as it predicts that the
$\T$-matrix vanishes logarithmically at $\omega=B$, a behavior that will
be corrected by including spin relaxation effects, see Sec.~\ref{sec:spin}.

\begin{figure}
\center
\includegraphics[width=0.30\textwidth]{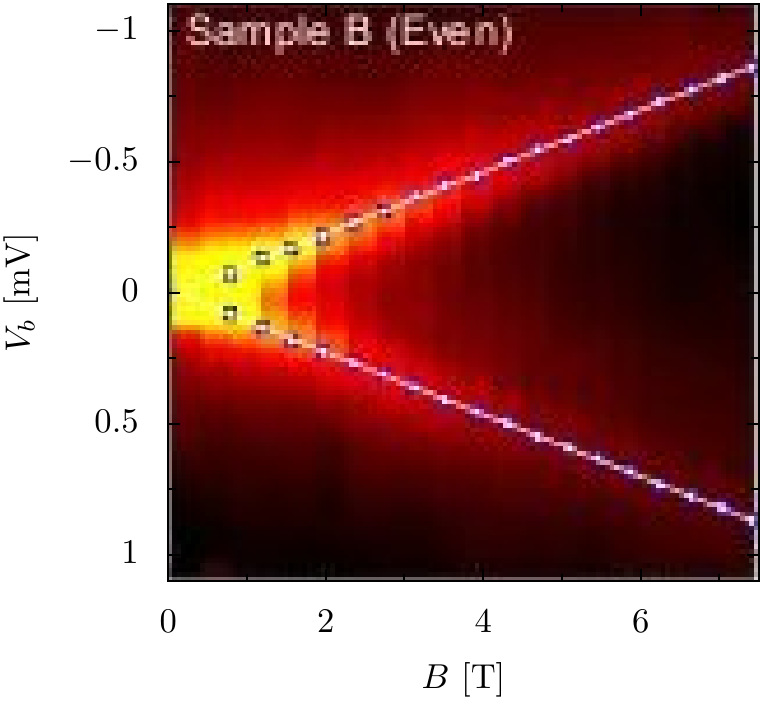}
\includegraphics[width=0.40\textwidth]{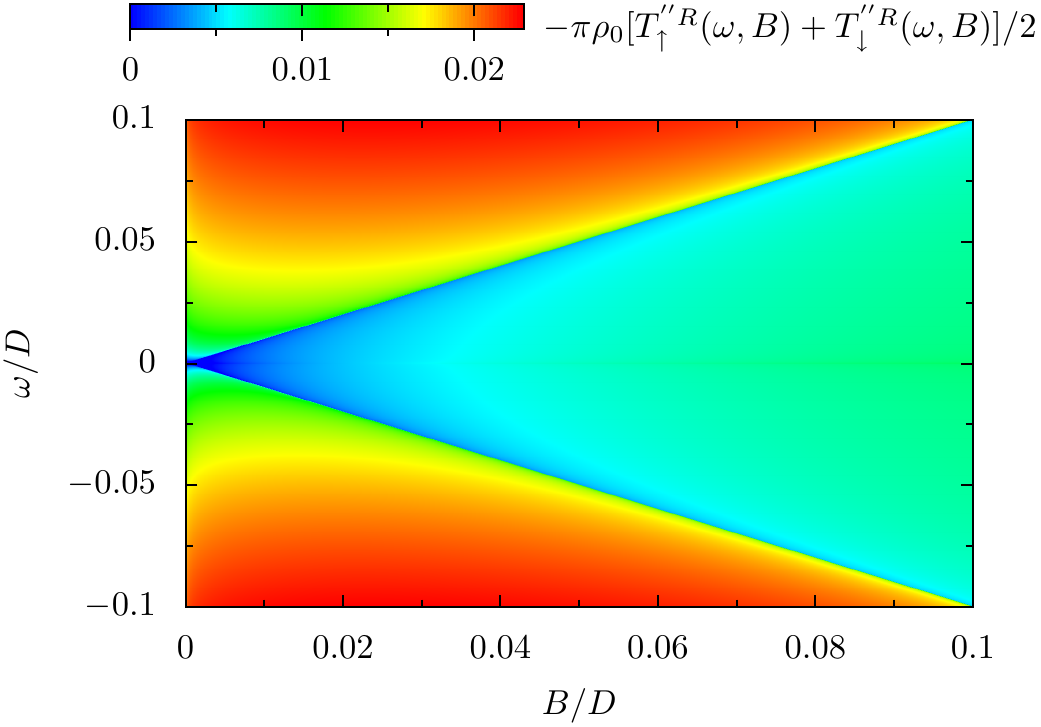}
\caption{\label{fig:ferro_splitting_experiment} (color online) Upper panel: experimental
data~\cite{roch_observation_2009} for the finite bias conductance of an 
underscreened $S=1$ Kondo dot as a function of magnetic field and applied bias. 
A splitting is still observed at small magnetic field, below the typical voltage scale characterizing 
the width of the $B=0$ differential conductance.
Lower panel: spin-symmetric $\T$-matrix of the $S=1/2$ Kondo model with
ferromagnetic exchange ($J/D=-0.138$) as a function of magnetic field and
frequency. Similarly to the experiment, the Zeeman effect persists even for 
vanishingly small magnetic field.}
\end{figure}

We finally address the issue of the Zeeman splitting in the spin-symmetric
$\T$-matrix, see Fig.~\ref{fig:ferro_splitting_experiment} which
compares experimental results for a $S=1$ underscreened molecular quantum
dot~\cite{roch_observation_2009} and our analytical result.
The analytical formula shows indeed that the Zeeman splitting persists for 
arbitrary small values of the applied magnetic field, giving strength to
the interpretation made in Ref.~\onlinecite{roch_observation_2009}.

\subsection{Spin relaxation time}
\label{sec:spin}

We finally investigate the role of spin relaxation in transport, motivated by
the spurious logarithmic singularities obtained at RPA level in the $\T$-matrix near 
the threshold $\omega=B$. 
Following previous
works~\cite{rosch_spectral_2003,rosch_kondo_2005,Kehrein,Schoeller,Schuricht,SchurichtAndergassen}, one can expect
that spin lifetime effects occuring at higher order in perturbation will smear
out the artificial divergence of the $\T$-matrix.
In the Majorana diagrammatics, spin relaxation is straightforwardly incorporated
within Majorana self-energies, because the magnetic susceptibilities identically
reduce to Majorana propagators, see {\it e.g.} Eq.~(\ref{newchiz}).

\begin{figure}[ht]
\includegraphics[width=0.47\textwidth]{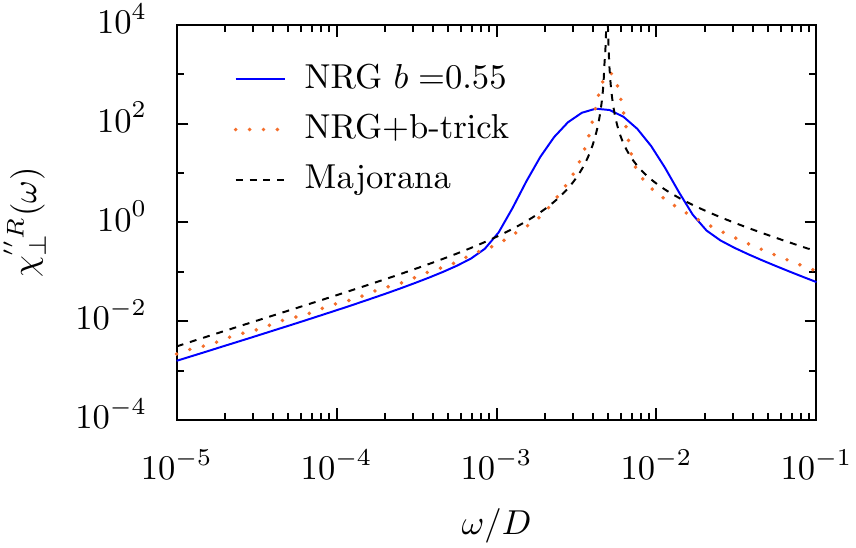}
\includegraphics[width=0.47\textwidth]{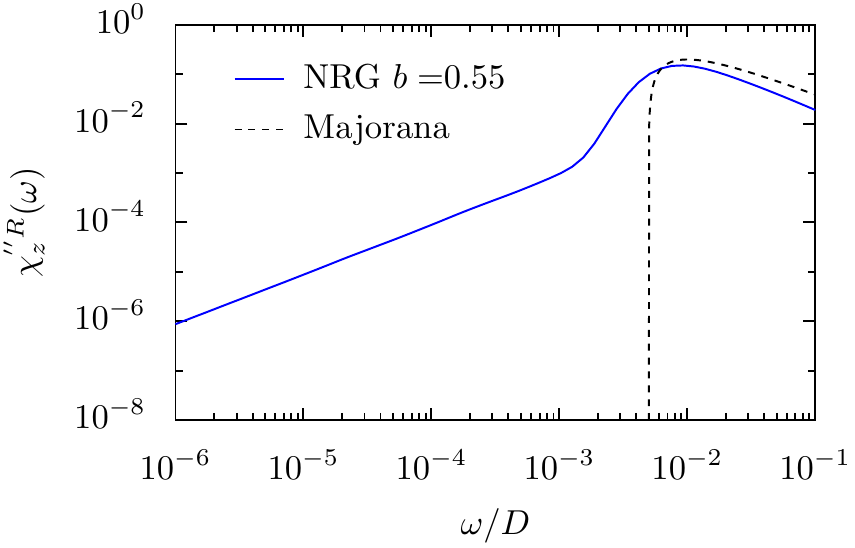}
\caption{\label{fig:self_plot_mag2} (color online)
Upper panel: transverse spin susceptibility for the antiferromagnetic Kondo
model, comparing the lowest order Majorana calculation to NRG simulations obtained 
with constant broadening~\cite{bulla_numerical_2008} $b=0.55$ and with the adaptative 
broadening methods (``b-trick'')~\cite{freyn_optimal_2009}, using the same parameters 
as in Fig.~\ref{fig:result_1}. 
The resonance at $\omega=B$ is clearly overbroadened, and the broadening trick
allows to improve on the resolution. 
Lower panel: longitudinal spin susceptibility within the Majorana diagrammatics
and the NRG (with constant broadening $b=0.55$ only). The spurious gap obtained
by the lowest order diagrammatics result should be filled by enforcing self-consistency in
the diagrammatic calculation scheme~\cite{Florens_spinboson}.}
\end{figure}

Due to the presence of the magnetic field along the $z$-axis, we need to introduce
longitudinal and transverse susceptibilities, 
$\chi_z(\tau)=-\left\langle T_{\tau}\left[S^z(\tau)S^z(0)\right]\right\rangle$ and 
$\chi_{\perp}(\tau) = -\left\langle T_{\tau} \left[ S^+(\tau)S^-(0)\right]\right\rangle$.
The spin operators are then replaced  by $S^{z}=\Phi\eta$,
$S^{+}=\sqrt 2 \Phi f^{\dagger}$ and $S^{-}=\sqrt 2 \Phi f$, so that imaginary parts of 
these susceptibilities are given by single fermion propagators:
\begin{eqnarray}
\label{eq:sus_equations2}
\chi^{''R}_z(\omega)&=-\frac{1}2\tanh\left(\frac{\beta\omega}{2}\right)G^{''R}_{\eta}(\omega)\\
\label{eq:sus_equations2b}
\chi^{''R}_{\perp}(\omega)&=-\tanh\left(\frac{\beta\omega}{2}\right)G^{''R}_f(\omega).
\end{eqnarray}
Spin relaxation is now immediately incorporated from Dyson's equation:
\begin{eqnarray}
\label{eq:Dyson1}
G_\eta^R(\omega)&=&\frac{1}{\omega-\Sigma_\eta(\omega)+i0^+}\\
\label{eq:Dyson2}
G_f^R(\omega)&=&\frac{1}{\omega-B-\Sigma_f(\omega)+i0^+}
\end{eqnarray}
and the needed imaginary part of the self-energies are easily computed up to 
second order in $J$ (here at zero temperature):
\begin{widetext}
\begin{eqnarray}\label{eq:self_energies1}
\Sigma_{\eta}^{''R}(\omega)&=& \delta \Sigma_\eta
-\frac{J^2}{4}\rho_0\int_{-D}^{D}\left[\Pi_{0f}^{''R}(\omega_1+\omega)-\Pi_{0f}^{''R}(-\omega_1-\omega)\right]
\left[\operatorname{Sign}(\omega_1+\omega)-
\operatorname{Sign}(\omega_1)\right] d\omega_1\\
\label{eq:self_energies2}
\Sigma_{f}^{''R}(\omega)&=&
-\frac{J^2}{4}\rho_0\int_{-D}^{D}\left[\Pi_{0f}^{''R}(\omega_1+\omega)+\Pi_{0\eta}^{''R}(\omega_1+\omega)\right]
\left[\operatorname{Sign}(\omega_1+\omega)-
\operatorname{Sign}(\omega_1)\right] d\omega_1,
\end{eqnarray}
\end{widetext}
from which the real part of the self-energies can be obtained by applying the
Kramers-Kronig relation. 
The self-energy $\Sigma_{\eta}$ contains an extra $\omega$-independent
contribution given by
\begin{equation}\label{eq:sigma_cf}
\delta\Sigma_{\eta}=-\frac{J^2}{4}\int_0^\beta d\tau G_{0c}(\tau)G_{0c}(-\tau) 
	= \frac{J^2}{4}\rho_0 \log(4),
\end{equation}
that originates from the last term in Eq.~(\ref{eq:Kondo_trafo}), and
which is crucial to preserve the correct particle-hole symmetry.
The susceptibilities obtained from the Majorana
propagators~(\ref{eq:sus_equations2})-(\ref{eq:sus_equations2b}) are shown in 
Figure~\ref{fig:self_plot_mag2}. 

Let us begin with the discussion of the transverse susceptibility, which is
expected to display a narrow resonance at the Zeeman energy. As noted 
previously~\cite{Florens_spinboson}, the NRG data for the transverse spin
susceptibility is much more overbroadened than the $\T$-matrix (which only 
shows a shoulder and not a peak), as can be gathered from the sharper feature 
found in the Majorana calculation. Some improvement of the NRG data processing
can be obtained by using adaptative broadening methods (``b-trick'')~\cite{freyn_optimal_2009}, 
where the broadening parameter $b$ is taken to be frequency-dependent. This allows 
to obtain a much narrower peak, yet still somewhat too broad. Larger scale
NRG simulations using so-called $z$-averaging~\cite{oliveira_generalized_1994} should 
allow to obtain better agreement.
We now turn to the longitudinal susceptibility, which similarly to the
$\T$-matrix is expected~\cite{Florens_spinboson} to give a shoulder at $\omega=B$.
The lowest order Majorana calculation shows good agreement with the NRG at
$\omega>B$, but fails to reproduce the low-frequency tails, because the
self-energy Eq.~(\ref{eq:self_energies2}) has a gap. Previous
work~\cite{Florens_spinboson} showed however that enforcing self-consistency in
the perturbative scheme allows to cure this defect of the bare perturbation
theory. Note also that the use of self-consistency in the self-energies allows to include
the renormalization of the exchange coupling within the relaxation rate, a small
quantitative effect that we did not take into account here.

\begin{figure}[ht]
\includegraphics[width=0.47\textwidth]{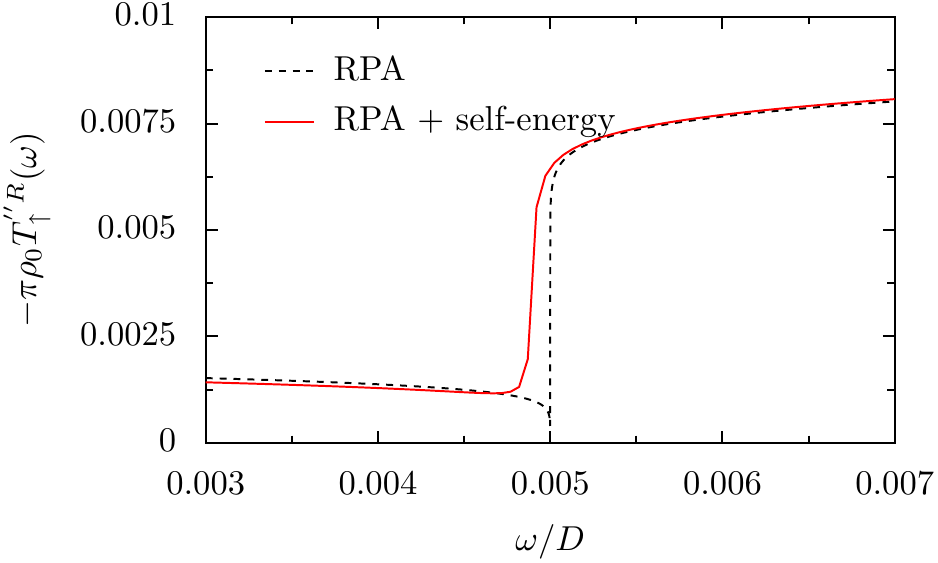}
\caption{\label{fig:T_changed} (color online) $\T$-matrix of the ferromagnetic Kondo model
near the Zeeman energy $B=0.005 D$, comparing the RPA calculation with and without
self-energy corrections. The inclusion of a finite spin relaxation rate
allows to cut the singularity (a spurious logarithmic vanishing at $\omega=B$)
of the bare Majorana calculation.}
\end{figure}

Ultimately, it is the finite amplitude of the resonance peak in the spin
susceptibilities at $\omega=B$ which is expected to regularize the spurious 
divergence in the
$\T$-matrix~\cite{rosch_spectral_2003,rosch_kondo_2005,Kehrein,Schoeller,Schuricht,SchurichtAndergassen}.
This is simply achieved by replacing the free propagators 
Eq.~(\ref{eq:Majorana_fermion_propag_trafo}) by the dressed ones
Eq.~(\ref{eq:Dyson1})-(\ref{eq:Dyson2})
into the previous $\T$-matrix calculation.
Figure~\ref{fig:T_changed} shows the inclusion of spin relaxation effects at
the RPA level for the $\T$-matrix for the ferromagnetic Kondo model.
The previously observed artifact, namely a logarithmic cancellation of $\T^{''R}$
at $\omega=B$, is now cured by the spin lifetime. Note still that the shoulder
at the Zeeman energy persists, an important check to argue about the occurrence
of a split Kondo resonance at low magnetic field in the underscreened
situation~\cite{roch_observation_2009}.

\section{Conclusion}

The goal of this paper was to understand quantum transport under an applied 
magnetic field in presence of a ferromagnetic Kondo interaction. Although magnetic 
impurities (or spin-active quantum dots) are most often coupled antiferromagnetically 
to conducting electrons (see however Ref.~\onlinecite{lucignano_kondo_2009}), the ferromagnetic 
case is relevant for the situation of underscreened large magnetic
moments~\cite{roch_observation_2009,ParksUS}. Indeed, partially screened moments 
decouple weakly from the electrons at low energy, due to an effective ferromagnetic 
Kondo interaction~\cite{nozieres_kondo_1980}. The existence of a degenerate
magnetic ground state leads then to a singular response to magnetic field, also
visible in transport, which we investigated here in some detail, using a
combination of density-matrix numerical renormalization group calculations and
an analytical perturbative technique based on Majorana diagrammatics~\cite{Florens_spinboson}.
This latter method is conceptually simpler than previously proposed various renormalization
scheme, either based on Callan-Symanzik equations~\cite{Fritz}, functional
RG~\cite{rosch_spectral_2003,rosch_kondo_2005}, flow equations~\cite{Kehrein},
and Liouvillian-based RG~\cite{Schoeller,Schuricht,SchurichtAndergassen}.
However we have to recognize that the Majorana diagrammatics is likely harder to 
implement beyond one-loop, because disentangling the sub-leading logarithmic
contributions really requires an RG method (the Majorana scheme performs yet very well beyond
one loop near a perturbatively-accessible quantum critical point~\cite{Florens_spinboson}).
Another issue addressed here and in previous
works~\cite{rosch_spectral_2003,rosch_kondo_2005} is the energy resolution of the
NRG, which is clearly insufficient. We believe that a careful comparison of
large scale NRG simulations~\cite{freyn_optimal_2009,Weichselbaum_Bfield,Zitko} 
and possibly two-loop RG analytics~\cite{Schoeller,Schuricht,SchurichtAndergassen},
already in the equilibrium situation, still need to be performed.

\begin{acknowledgments}

We wish to thank J. Paaske, A. Shnirman, and G. Zar\'and for useful discussions,
A. Freyn for many helpful advices, and the authors of the flexible DM-NRG code
({\tt http://www.phy.bme.hu/$\sim$dmnrg/}). 
Support from ERC Advanced Grant MolNanoSpin n°226558 is also acknowledged.

\end{acknowledgments}

\end{document}